# Title: A transmission electron microscopy study of presolar hibonite

# Running head: TEM of presolar hibonite

Thomas J. Zega[1*], Conel M. O'D. Alexander[2], Larry R. Nittler[2], and Rhonda M. Stroud[1]

[1]Materials Science and Technology Division
Naval Research Laboratory
4555 Overlook Ave SW, Washington D.C. 20375
*corresponding author (tzega@nrl.navy.mil)

[2]Department of Terrestrial Magnetism
Carnegie Institution of Washington
5241 Broad Branch Rd NW
Washington D.C. 20015

**Abstract**

We report isotopic and microstructural data on five presolar hibonite grains (KH1, KH2, KH6, KH15, and KH21) identified in an acid residue of the Krymka LL3.1 ordinary chondrite. Isotopic measurements by secondary ion mass spectrometry (SIMS) verified a presolar circumstellar origin for the grains. Transmission electron microscopy (TEM) examination of the crystal structure and chemistry of the grains was enabled by *in situ* sectioning and lift-out with a focused-ion-beam scanning-electron microscope (FIB-SEM). Comparisons of isotopic compositions with models indicate that four of the five grains formed in low-mass stars that evolved through the red-giant/asymptotic-giant branches, whereas one grain formed in the ejecta of a Type II supernova. Selected-area electron-diffraction patterns show that all grains are single crystals of hibonite. Some grains contain minor structural perturbations (stacking faults) and small spreads in orientation that can be attributed to a combination of growth defects and mechanical processing by grain-grain collisions. The similar structure of the supernova grain to those from RGB/AGB stars indicates a similarity in the formation conditions. Radiation damage (e.g. point defects), if present, occurs below our detection limit. Of the five grains we studied, only one has the pure hibonite composition of $CaAl_{12}O_{19}$. All others contain minor amounts of Mg, Si, Ti, and Fe. The microstructural data are generally consistent with theoretical predictions, which constrain the circumstellar condensation temperature to a range of 1480 K to 1743 K, assuming a corresponding total gas pressure between $1 \times 10^{-3}$ and $1 \times 10^{-6}$ atm. The TEM data were also used to develop a calibration for SIMS determination of Ti contents in oxide grains. Grains with extreme $^{18}O$ depletions, indicating deep mixing has occurred in their parent AGB stars, are slightly Ti-enriched compared to grains from stars without deep mixing, most likely reflecting differences in grain condensation conditions.

**Introduction**

As stars evolve, they shed matter through dust-driven stellar winds or explosive events such as supernovae. These stellar ashes enter the interstellar medium and become the starting material for new stars. Our own solar system is believed to have partly formed from the remnants of ancient stars, and it was long ago suspected that individual grains of this presolar stardust material should have survived intact within the solid relics leftover from its birth, i.e., primitive meteorites (Boato 1954; Reynolds & Turner 1964; Black & Pepin 1969).

The isolation and extraction of presolar grains has been a decades-long struggle, partially motivated by finding the carrier phases for isotopically anomalous Ne and Xe in some primitive meteorites (see Anders & Zinner 1993 for a review). Early workers showed that physical separation and acid-dissolution treatments were effective at removing the bulk of the material contained within primitive meteorites, which largely formed within our solar system, and isolating the presolar fraction (Lewis et al. 1987). Nanodiamond, SiC, and graphite were among the first presolar phases to be recognized (Bernatowicz et al. 1987a; Lewis, et al. 1987; Tang & Anders 1988; Amari et al. 1990), and since these early efforts, presolar silicates, carbides, metal, nitrides, and oxides have been identified with various techniques (Bernatowicz et al. 1987b; Croat et al. 2003; Daulton et al. 2003; Nguyen & Zinner 2004; Stadermann et al. 2005; Messenger et al. 2005; Stroud et al. 2004, 2006; Vollmer et al. 2007; Floss et al. 2008; Nittler et al. 2008; Zega et al. 2009).

Much of what we know about presolar grains has come from detailed measurements of their isotopic compositions. When compared with astrophysical models and remote observations, isotopic measurements have provided a wealth of information on the types, compositions, and masses of stars from which the grains originated (Nittler 1997; Clayton & Nittler 2004; Zinner 2005). Detailed information on crystal structure and chemistry can also provide important information on the history of presolar grains, e.g., thermodynamics of circumstellar envelopes, physical processing in the interstellar medium, and the effects of meteorite parent-body processes. For example, the microstructural properties of presolar graphite and SiC grains have received considerable attention. Bernatowicz et al. (1996) used TEM to examine the microstructure of hundreds of graphite spherules from the Murchison (CM2) chondrite from which they inferred formation pressures and C number densities of the circumstellar envelopes in which the grains formed. In their exhaustive study of SiC, Daulton et al. (2002; 2003) used TEM to examine 508 individual grains from Murchison and found that the vast majority of them (82.1%) occur as the 3C and 2H polytypes. They concluded that the SiC formed as these two polytypes because the low pressures in circumstellar envelopes result in low condensation temperature for SiC. Croat et al. (2005) examined 847 presolar graphite grains, also from Murchison, and inferred from large *s*-process element enrichments that most of them formed in the outflows of asymptotic giant branch (AGB) stars. Modeling by Bernatowicz et al. (2005) put further constraints on the conditions of graphite formation on the basis of the TEM data. It is important to note that the size (up to several μm in diameter) of many of the grains in the above studies permits sample preparation by conventional methods (e.g. ultramicrotoming) and facilitates analysis of large numbers of them. Moreover, in the meteoritic acid residues, essentially all SiC and graphite grains have been found to be presolar, eliminating the need to establish their origin by first making isotopic measurements. In comparison, only a small fraction of oxide grains, e.g., hibonite ($CaAl_{12}O_{19}$), in residues are actually presolar (Choi et al. 1999; Nittler et al. 2008), which combined with their small grain sizes (μm down to hundreds of nm), has made it particularly challenging to acquire both isotopic and microstructural data on the same grains.

In the past decade, several developments have been key to presolar-grain studies. Automated mapping techniques for secondary ion mass spectrometry or SIMS (Nittler & Alexander 2003; Gyngard et al. 2010) have greatly enhanced our ability to identify efficiently and measure rare types of presolar grains. The advent of the NanoSIMS ion microprobe, with its smaller probe size than previous generations of instruments, furthered the automated mapping capabilities by permitting detection of grains down to ≈100 nm in size, with sufficient precision and accuracy to identify presolar materials and provide useful constraints on their astrophysical origins (Zinner et al. 2003; Stadermann et al. 2005). The other key development for microstructural studies has been the focused-ion-beam scanning-electron microscope (FIB-SEM) and *in situ* lift-out capabilities. The FIB-SEM combines the non-destructive imaging capabilities of the field-emission SEM with the sputtering capabilities of a 10-nm ion beam. Once presolar material is identified using SIMS techniques, the FIB-SEM can be used to precisely section, extract, and thin a presolar grain (≈100 nm for electron transparency) for detailed crystallographic investigation by transmission electron microscopy (TEM). The coordinated use of SIMS, FIB-SEM, and TEM is a powerful combination for acquiring isotopic and crystallographic information from the same presolar grain (Stroud et al. 2004; Zega et al. 2007). Here for the first time we apply the coordinated approach to the analysis of presolar hibonite grains.

Hibonite is a member of the magnetoplumbite mineral group and in pure form its composition is $CaAl_{12}O_{19}$. It consists of a layered structure (space group $P6_3/mmc$, a = 0.556 nm and c = 2.19 nm) with a close-packed oxygen sublattice perpendicular to the **c** axis. Cations occur interstitial to the O framework in octahedral, tetrahedral, and trigonal bypyramidal sites (see Bermanec et al. 1996 and Hofmeister et al. 2004 for a more detailed description). Although relatively rare in terrestrial formations, hibonite has been reported in granulite facies metamorphic rocks (e.g. Rakotondrazafy et al. 1996; Sandiford & Santosh 1991), and while predominantly a calcium aluminate, it can contain Mg, Ti, Fe, and Si with minor amounts of La, Ce, Nd, and Th (Bermanec et al. 1996; Maaskant et al. 1980). In meteorites, hibonite occurs as micrometer-sized lathic grains in calcium-aluminum-rich inclusions, and can contain Mg, Ti, Fe, V, and minor Si (Simon et al. 2006 and references therein). Equilibrium thermodynamic calculations predict hibonite to be the second major oxide to condense from a gas of solar composition (Lodders 2003), and comparison of laboratory-based and remotely sensed infrared spectra indicate that it might occur in planetary nebulae (Hofmeister et al. 2004). Knowledge of the structure and composition of presolar hibonite can, therefore, provide a basis for comparison with its solar and terrestrial counterparts, insight into circumstellar processes, and ground truth for astronomical observations. We report here a study of five presolar hibonite grains from the Krymka unequilibrated ordinary chondrite (LL3.1). The isotopic characteristics of the grains are reported and discussed in detail by Nittler et al. (2008), and so here we focus on their chemical and structural properties.

**Methods**

The studied presolar hibonite grains studied here were identified in an acid residue of the Krymka LL3.1 ordinary chondrite by use of an automated particle isotopic analysis system (Nittler & Alexander 2003). Of some 7,000 measured micron-sized oxide grains from the residue, 21 were determined to be presolar hibonite on the basis of unusual O-isotopic ratios. Follow-up NanoSIMS measurements provided further information on the isotopic composition of Mg, K, and Ca in many of the grains. Five of these presolar hibonites were selected for TEM analysis. Details of the Krymka residue preparation and isotopic measurements are given by Nittler et al. (2008).

We used an FEI Nova 600 focused-ion-beam scanning-electron-microscope (FIB-SEM) at the Naval Research Laboratory to make electron-transparent sections of five presolar hibonite grains from the Krymka LL3.1 ordinary chondrite. The Krymka hibonite (KH) grains examined in this study are hereinafter referred to as KH1, KH2, KH6, KH15, and KH21. All grains were lifted out *in situ* using FIB methods similar to those described by Zega et al. (2007) with the exception that grain KH2 was welded to a Mo grid rather than extracted using a microtweezer.

FIB sections of grains KH1, KH2, KH15, and KH21 were examined with a 200 keV JEOL 2200FS TEM equipped with an energy-dispersive X-ray spectrometer (EDS) and bright- and dark-field scanning TEM (STEM) detectors. Grain compositions were determined with an ultra-thin-window Thermo Electron EDS detector and processed with Noran System Six software. Depending upon illumination conditions, the spectral acquisition time was varied between one and five minutes to ensure good counting statistics (high count rate with spectrometer dead time ≤30%). All spectra were fitted with a Gaussian model and quantified based on detector-sensitivity (k) factors (Cliff & Lorimer 1975) derived from standards. San Carlos olivine and Hakone anorthite were used to calculate k factors for Mg and Si, whereas a FIB section of a terrestrial sample of hibonite from the Furua Granulite Complex in southern Tanzania (see Maaskant et al., 1980, for details on its composition) was used for Ca, Al, and Ti. The FIB section of grain KH6 was characterized with a 200 keV JEOL 2010F TEM equipped with a Noran Vantage ultra-thin window EDS system. This system used the same detector as that for the other grains in this study, and the counting statistics were consistent across the measurements. Thus, both systems have similar detection limits. Standardless quantification with Cliff-Lorimer-type refinement, including absorption correction, was used to obtain the elemental composition of this grain. The FIB section detached from the microtweezer support and was lost, preventing further EDS analysis with standards.

Selected-area electron-diffraction (SAED) patterns were acquired, where possible, from multiple grain orientations. All SAED patterns were measured, both manually (using Adobe Photoshop) and with the crystallographic image processing software package, CRISP (Hovmoller 1992), based on calibrated camera constants. The indexing of the SAED patterns was based on hibonite lattice parameters and symmetry, and verified by comparison to diffraction patterns calculated using JEMS multislice simulation software (Stadelmann 1987).

**Results**

The O- and Mg-Al-isotopic compositions of the hibonite grains studied here are given in Table 1 (additional Ca-isotopic data for three of the grains can be found in Nittler et al. 2008). The O isotopic compositions of presolar hibonite grains span a range comparable to that of previously studied presolar $Al_2O_3$ and $MgAl_2O_4$ (Fig. 1), and those chosen for this study plot within the Group 1 (grains KH6 and

143   KH21), Group 2 (KH1 and KH15), and Group 4 (KH2) fields for presolar oxide grains (Fig. 1) as defined by

144   (Nittler 1997). The Group locations, indicated by broad ellipses in three-O isotope space (Fig. 1), are

145   approximate with some overlap among them. Nonetheless, the Groups highlight trends that reflect the

146   properties of their parent stars, and we will discuss these below.

147         The hibonite grains, as imaged after SIMS analysis, range in size from hundreds of nanometers

148   to several microns measured by orthogonal dimensions in secondary electron images (Fig. 2). Each of

149   the samples consist of a single hibonite grain sitting on top of a Au pedestal, produced by the differential

150   sputtering of Au and hibonite during SIMS analysis, except that of KH6 (Fig. 2a), which consists of a

151   hibonite grain sitting on top of an isotopically normal $Al_2O_3$ grain (cf., Fig. 2a,b). The spatial association

152   of KH6 with the underlying $Al_2O_3$ is an artifact of sample preparation and not a result of growth or

153   reaction during circumstellar condensation. The surface topologies of the grains vary from smooth and

154   flat (e.g. KH21, Fig. 1c) to high relief from the underlying Au pedestal (e.g. KH6, Fig. 1a). We present the

155   TEM data for each of grains separately below.

156   KH6

157         Group 1 grain KH6 measures 153 nm x 322 nm and occurs above a 560 nm x 1000 nm corundum

158   grain, and both are sandwiched between the conductive Au substrate and the Pt strap deposited during

159   the FIB-SEM preparation (Fig. 3a). There is no observable variation in diffraction contrast within the

160   grain, and the SAED patterns acquired from the hibonite shows that it is a single crystal (Fig. 3b).

161   Reflections within the SAED pattern reveal intensity variations, with some exhibiting satellite spots

162   indicative of microtwins or stacking disorder. Standardless quantification of EDS spectra from this grain

163   (Table 1, KH6) is consistent with a nominally stoichiometric $CaAl_{12}O_{19}$ composition; no Mg, Si, Ti, or Fe

164   were observed down to our detection limit (≈0.1 at%).

165

166    KH21

167        Group 1 grain KH21 is approximately 3.7-μm long and its width varies between 320 and 775 nm

168    (left- and right-hand sides, respectively) as shown in the BF and high-angle annular-dark-field (HAADF)

169    images of the FIB section (Fig. 4a,b). The BF image reveals that most of the grain has uniform diffraction

170    contrast except for the right-hand side, which exhibits horizontal striations that parallel the bottom edge

171    of the grain and extend for approximately 1 μm (Fig. 4a). An amorphous damage layer ranging from 10

172    to 15 nm thick extends along the top surface of the grain as a result of the SIMS analysis. The SAED

173    pattern for the [1100] zone axis shows streaking along c* of the hibonite structure (Fig. 4c), and the high

174    resolution TEM (HRTEM) image reveals abundant stacking disorder along [0001] (Fig. 4d). The HAADF

175    image contains mostly uniform contrast except for the right-most side of the grain where it is slightly

176    mottled in and around the area exhibiting stacking disorder. EDS gives an average composition,

177    calculated assuming 19 O atoms, of $Ca_{1.01}Al_{11.73}Mg_{0.21}Ti_{0.07}Si_{0.01}Fe_{0.01}O_{19}$ for the bulk of the grain. The area

178    containing the stacking disorder is depleted in Ca (10.2%) and Mg (9.7%) but slightly enriched in Al

179    (0.8%) relative to the average bulk composition.

180

181    KH15

182        Group 2 grain KH15 exhibits an atypical morphology relative to the other grains in this study. It

183    contains a central region that measures approximately 2.4-μm wide by 1.4-μm high and a segment that

184    branches off the top-right corner and extends upward to the left-side of the FIB section (we refer

185    interested readers to Fig. 10 of Zega et al., 2007, for images of this grain acquired during FIB-SEM

186    preparation). The segment measures 2.5-μm long and its width varies between 120 and 360 nm (Fig.

187    5a,b). Bright-field imaging does not reveal a grain boundary between the central region and the segment

188    that extends above it (Fig. 5a). Zone-axis SAED patterns acquired from several areas suggest an angular

189    variation ranging from 5.1° to 14.1° between the central part of the grain and the segment above (Fig.

5c-e). The HAADF image shows uniform contrast for the entire crystal (Fig. 5b), suggesting a homogeneous composition. Quantification of EDS spectra give an average composition of $Ca_{1.06}Al_{11.69}Mg_{0.06}Ti_{0.17}Fe_{0.01}O_{19}$.

KH1

Group 2 grain KH1 measures 1.34 μm across and exhibits variable heights ranging from 50 nm at the left edge up to 200 nm in the center (Fig. 6a). Bright-field imaging shows uniform contrast throughout most of the grain except the right-most 300 nm where there is some variation in the diffraction contrast likely due to stacking disorder. The HAADF image shows that the grain has uniform contrast throughout (Fig. 6b), suggesting a homogeneous composition. The SAED pattern acquired from the grain indicates that it is also a single crystal (Fig. 6c). EDS quantification gives a composition of $Ca_{1.04}Al_{11.58}Si_{0.09}Mg_{0.14}Ti_{0.11}Fe_{0.06}O_{19}$.

KH2

Measurements from the bright-field TEM image show that Group 4 ($^{18}$O-rich, Fig. 1) grain KH2 is 600-nm wide by 230-nm high, with a crack extending 200 nm into the grain from the left side (Fig. 7a,b). The top 30 nm of the grain does not exhibit diffraction contrast and is amorphous, most likely due to ion beam damage during the SIMS measurements. No evidence for damage from either cosmic or laboratory processing is seen on the underside of the grain. The HAADF image (Fig. 7b) shows that the grain is mostly uniform except for the top amorphous region where it appears slightly darker, indicative of a decrease in density, perhaps due to implantation of $^{16}$O from the SIMS measurements (Fig. 7b). The surface damage of a grain depends on several variables including, e.g., the initial grain geometry prior to SIMS, the conditions used in the SIMS analysis, and the final thickness of the grain after FIBing. For some or all of these reasons the damage layer in KH2 is readily observed, whereas in other grains, e.g., the top surface of KH15 (Fig. 5), it is less pronounced. The remainder of the grain is crystalline and its diffraction patterns index to a single hibonite crystal (Fig. 7c), but the crystal orientation varies by a few degrees across the grain. Quantification of the EDS spectra gives a formula of $Ca_{0.98}Al_{11.77}Si_{0.02}Mg_{0.14}Ti_{0.09}Fe_{0.01}O_{19}$.

**Ti contents of presolar oxide grains**

For three of the hibonite grains analyzed here, we had previously obtained $^{48}Ti^+/^{27}Al^+$ secondary ion ratios by NanoSIMS measurements. Based on the quantitative EDS measurements of these three grains, we derived a relationship between Ti contents and secondary ion ratios:

$$[Ti] (wt. \%) = 0.078 + 149.3*(^{48}Ti^+/^{27}Al^+)$$

This formula was also found to reproduce TEM-EDS Ti abundances for two presolar $Al_2O_3$ grains (Stroud et al. 2004) within ≈40%, so we believe that absolute Ti contents derived in this way are accurate to this level for both hibonite and $Al_2O_3$ grains. Relative abundances are probably accurate to ≈25%. Table 2 gives Ti abundances for 31 presolar grains, five determined by TEM-EDS and the remainder from the SIMS secondary ion ratios. Figure 8 shows Ti abundances versus $^{18}O/^{16}O$ ratios for the presolar oxide grains. Two general features are clear: first, hibonite grains have higher Ti contents than $Al_2O_3$ grains (average hibonite = 0.59 wt%, average $Al_2O_3$ = 0.17 wt %) and second, $^{18}O$-depleted (Group 2) grains typically have higher Ti contents than do Group 1 grains. All Group 2 hibonite grains contain more Ti than the most Ti-rich Group 1 hibonite grains. There is more scatter for $Al_2O_3$ grains, but three of six Group 2 grains have higher Ti than the most Ti-rich Group 1 $Al_2O_3$ grain.

**Discussion**

The isotopic compositions of presolar grains reflect both the initial compositions of the parent stars, determined by Galactic Chemical Evolution (GCE) processes, and the nuclear processing and mixing that occurred in the parent stars themselves. Nittler and co-workers (Nittler 1997; Nittler et al.

1994) showed that most presolar O-rich grains tend to plot within four distinct groups in three-O isotope space. Group 1 grains are believed to originate in low-mass (≈1.2 to 2.5 $M_\odot$) red giant branch (RGB) and AGB stars. Comparisons of their O isotopic ratios with model predictions for such stars (e.g. Boothroyd & Sackmann 1999) can be used to estimate the masses and metallicities of their parent stars. In general, the $^{17}O/^{16}O$ ratio is sensitive to stellar mass and the $^{18}O/^{16}O$ ratio to metallicity. For example, based on its O isotopes, grain KH6 is inferred to have formed in a solar-metallicity RGB or AGB star of roughly 1.4 $M_\odot$, and the O and Mg isotopes of KH21 (including a strong $^{25}Mg$ depletion) indicate an origin in a ≈1.5$M_\odot$ AGB star with metallicity ≈0.75 times solar (Nittler et al. 2008).

The enhanced $^{17}O/^{16}O$ ratios of Group 2 grains (e.g., KH1 and KH15) suggest an origin in low-mass (<1.8$M_\odot$, Nittler et al. 2008) AGB stars. However, the large $^{18}O$ depletions and initial $^{26}Al/^{27}Al$ ratios (inferred from $^{26}Mg$ enrichments) observed in these grains are not predicted by standard stellar evolution models and point to the need for an extra-mixing process, called cool bottom processing, to have occurred in the parent stars (Nollett et al. 2003; Wasserburg et al. 1995). Unfortunately, the metallicity cannot be independently inferred for the parent stars of Group 2 grains because their compositions before the extra mixing is unknown. In cool bottom processing $^{18}O$ is destroyed via $^{18}O(p,\alpha)^{15}N$, and therefore standard models cannot be used to infer initial compositions. However, it has been argued that cool bottom processing should be more efficient in stars with lower-than-solar metallicity (Boothroyd & Sackmann 1999).

The origin of the $^{18}O$-enriched Group 4 grains has been more enigmatic, with high-metallicity stars, unusual AGB stars, and supernovae all having been suggested (Choi et al. 1998; Nittler et al. 1997). However, as discussed in detail by Nittler et al. (2008), multi-element isotope data for two Group-4 oxide grains, including KH2, clearly point to an origin in the ejecta of Type II supernovae for these grains, and by extension for most or all Group-4 grains. Similar support for a supernova origin of Group-4 grains was recently provided by Mg isotopic measurements of $^{18}$O-enriched presolar silicates (Nguyen et al. 2010).

Despite origins in such different stellar environments discussed above, the TEM data for each of the grains are broadly similar. All SAED patterns show that the grains are single crystals and have stoichiometric compositions. Of the five grains examined in this study, only one (KH6) is pure $CaAl_{12}O_{19}$, whereas the others contain minor amounts of Mg, Ti, Si, and Fe. Thus, it appears that all grains condensed in near thermodynamic equilibrium with the ambient gas, which can be used to constrain the circumstellar condensation temperature and pressure conditions. Equilibrium calculations generally predict that hibonite ($CaAl_{12}O_{19}$) is the first phase after corundum ($Al_2O_3$) to condense from a cooling gas of solar composition (Yoneda & Grossman 1995; Ebel & Grossman 2000; Lodders 2003). The estimated condensation temperature ($T_c$) depends on the assumed total gas pressure ($P_T$), ranging from 1743 K at $10^{-3}$ atm to 1480K at $10^{-6}$ atm (Yoneda & Grossman 1995; Ebel & Grossman 2000; Lodders 2003). As $T_c$ decreases with decreasing total pressure, 1743 K might represent an upper temperature limit assuming that the $P_T$ in the circumstellar environment from which these grains condensed was $\leq 1 \times 10^{-3}$ atm. This assumption is reasonable given that models estimate pressures in the photospheric region of C stars to range from $10^{-3}$ to $10^{-5}$ atm, with lower pressures expected in the envelope (Lodders & Fegley 1995).

In principle, the grains' condensation conditions could be further constrained using the inferred metallicities of their parent stars and the measured abundances of substitutional impurities in the grains. The slope of the stability field for hibonite at solar metallicity ($P_T \approx 10^{-3}$ atm) suggests that its condensation temperature decreases with decreasing metallicity (e.g. see plate 10 of Ebel 2006). Moreover, incorporation of impurities, such as Ti, Si, Mg, and Fe is likely to increase with decreasing temperature because the sticking efficiency for impurities incorporated onto grain surfaces generally increases at lower temperature. Thus, grain KH21, which comes from a sub-solar metallicity star and has a higher impurity concentration, may have condensed at a lower temperature than grain KH6, which had no detectable impurities and comes from a solar metallicity star. However, the metallicity is only available for these two of the five grains, and thus no strong correlation between metallicity and impurity concentration can be determined. Moreover, the lack of correlation between the $^{18}O/^{16}O$ ratio and Ti contents for a larger number of Group 1 grains (Figure 8) also argues against a correlation between metallicity and impurity concentrations. Further quantitative assessment of the formation conditions of grains KH1, KH2, and KH15 is difficult without thermodynamic data on the solid solution of Ti, Si, Mg, and Fe into hibonite, which, to our knowledge, are unknown. Nonetheless, given the range of impurities and their concentrations in solar-system hibonites, e.g., ≤1.11 wt% FeO, ≤1.39 wt% $SiO_2$, ≤4.22 wt% MgO, 0.14 to 8.73 wt% $TiO_2$, ≤1.8 wt% $V_2O_3$, ≤0.13 wt% $Cr_2O_3$, ≤0.04 wt% MnO (Keil and Fuchs, 1971; Armstrong et al. 1982; Michel-Lévy et al. 1982; El Goresy et al. 1984; MacPherson and Grossman, 1984; Beckett et al. 1988; Fahey et al. 1994; MacPherson and Davis, 1994; Simon et al. 1994; Greenwood et al. 1994; Bischoff and Srinivasan, 2003; Simon et al. 2006), the variations in Fe ,Si, Mg, and Ti contents of the presolar grains (0.10 to 0.67 wt% FeO, 0.13 to 0.78 $SiO_2$, 0.33 to 1.27 wt% MgO, and 0.86 to 2.04 wt% $TiO_2$, Table 1) are mostly attributable to temperature and pressure variations in the individual circumstellar envelopes.

Particularly noteworthy is that the condensation conditions for the supernova grain (KH2) do not appear to be significantly different from that of the other grains studied (KH2, KH6, KH25, and KH21), which condensed in RGB/AGB stars. This is in strong contrast to the case of presolar SiC, where AGB-derived grains are typically single crystals (Daulton et al. 2003) but supernova grains are polycrystalline aggregates (Stroud et al. 2005; Hynes et al. 2010). One clue to the difference might lie in the different portions of the supernova ejecta implicated in the different types of grains. Whereas supernova SiC grains likely condensed in the deep C-rich layer that has experienced partial He-burning, with some contribution from even deeper zones, the composition of grain KH2 is best explained as consisting of ≈93% H-rich envelope material with the remainder coming from interior zones. Our data suggest that there are regions within the envelope of supernovae that have P-T conditions similar to those of RGB/AGB stars that are suitable for condensing single crystals of hibonite, whereas the condensation conditions for dust forming deeper in the ejecta are different. The polycrystalline nature of a supernova olivine [(Mg,Fe)$_2$SiO$_4$)] reported by Messenger et al (2005) is interesting in this regard, as its composition is also dominated by material from the He-C rich layer.

As seen in Figure 8, Group 2 hibonite (and perhaps Al$_2$O$_3$) grains appear to systematically have slightly higher Ti contents than Group 1 grains. The origin of this difference is unclear. One possibility is that the Group 2 parent stars had higher Ti abundances, relative to Al, than did the Group 1 parents. However, there is no obvious explanation for such abundance differences among low-mass AGB stars. For example, stellar observations indicate that the Ti/Al ratio of stars is roughly independent of metallicity (Reddy et al. 2006). Moreover, the cool bottom (extra-mixing) processes invoked to explain the low $^{18}$O/$^{16}$O ratios of Group 2 grains would not be expected to affect Ti or Al abundances, at least on nucleosynthesis grounds. The physical mechanism responsible for cool bottom processing in AGB stars is not well understood, but one recent suggestion is magnetic buoyancy induced by stellar dynamos (Busso et al. 2007; Nordhaus et al. 2008). In the case of the Sun, magnetic phenomena induce large elemental fractionations in the corona, raising the intriguing possibility that similar effects could play a role in explaining the grain data. Perhaps more likely, the difference in Ti contents is related to differences in the crystal structures of the grains and the physical conditions of grain formation in the stars, for example temperature and/or pressure. Unfortunately, without a quantitative understanding of the parameters influencing Ti (and other minor element) contents during hibonite condensation, it is not possible to come to further conclusions.

Despite being single crystals, the grains exhibit minor structural variations that reflect a mixture of growth and processing effects. For example, SAED patterns from three of the grains (KH1, KH6, and KH21) show evidence for stacking disorder, which could result from either slightly non-equilibrium growth conditions, or subsequent mechanical processing, i.e., shear-transformation due to grain-grain collisions in the ISM or solar nebula. The large crack running through the center of the grain KH2 (Fig. 7a,b) and the spread in crystallographic orientation across it is most likely the result of grain fracture during a collision event, either in the SN outflow, the ISM, or the solar nebula. The morphology of grain KH15 is distinct from the other four grains in that it contains an arm that extends at an angle to the bulk of the grain (Fig. 5). There is no resolvable grain boundary between the bulk of the crystal and the arm despite the small angular spread between these two regions. Thus, these do not appear to be separate crystals that came together in the circumstellar envelope of the parent star. Rather, we infer that the angular spread between these regions is the result of changes in the growth direction during condensation, possibly due to decreasing temperature. The morphologies of grains KH1 and KH21 are similar to the arm of grain KH15, and they share the additional common feature of high defect densities on their thickest ends (Figs. 4,6). It is possible that grains KH1 and KH21 are fragments of larger condensates with complex platy morphologies that were similar to KH15, and the defects are concentrated near the point of fracture between the observed grain and other plates.

There is no significant evidence for radiation-induced processing of the grains, other than that from the SIMS measurements mentioned above, either in space or during laboratory processing. The specific susceptibility of hibonite to radiation damage is to our knowledge unknown. However, it is likely to be similar to other ionically bonded Al-rich oxides, such as corundum and spinel, which are relatively robust against radiation-induced amorphization, and have been studied in synthetic form for use as reactor materials (Clinard et al. 1982; Hobbs et al. 1994; McHargue 1987; Wang et al. 1998; Zinkle & Pells 1998). Radiation processing has not been detected in either presolar corundum (Stroud et al. 2004) or spinel grains (Zega et al. 2008; 2009; 2010). However, such signatures have been observed as amorphous coatings on TiC and kamacite grains within presolar graphites from supernovae (Croat et al. 2003) and as amorphous surface coatings on and tracks within olivine grains in interplanetary dust particles (Bradley et al. 1984). Furthermore, the effects of radiation processing on some dust grains can be severe: <1% of interstellar silicates survive radiation-induced amorphization (Kemper et al. 2004). Protection against radiation damage in the ISM by mantles on the grain surfaces (Nuth et al. 2000), could explain the lack of detectable radiation damage in the oxide grains, however it is unlikely that such mantles would selectively protect oxides and not silicates. Thus, it is more likely that some radiation damage of the oxides does occur, but primarily in the form of isolated point defects (Hobbs et al. 1994), which are not directly observable, even in the HRTEM images of the FIB slices. This is consistent with prior laboratory radiation damage studies in which clustering of the defects occurs only at temperatures of hundreds of K. For example, Zinkle & Pells (1998) subjected samples of $Al_2O_3$ to 4 MeV $Ar^+$ ions at 200 and 300 K and doses of 0.1 to 10 displacements per atom. They did not observe amorphization at either temperature and found dislocation loops and network dislocations formed at 300 K, whereas no defect clusters formed at 200 K. In comparison, Clinard et al. (1982) report the formation of defect clusters under neutron irradiation at 400 K which anneal at 900 K into interstitial dislocation loops lying on (10-10) planes.

Planar condensation of interstitials onto a new set of crystallographic sites is a common aggregation mode for defect formation in irradiated crystals and can give rise to the formation of stacking faults. For example, Howitt & Mitchell (1981) showed that corundum ($\alpha$-$Al_2O_3$), an ionic insulator related to hibonite, can condense interstitials onto the (0001) or (01-10) planes faulting the Al-cation sublattice rather than that of the O. The stacking sequence along either direction can be regenerated and the stoichiometry maintained with a partial shear across the plane of the dislocation loop. However, such planar aggregation occurs at high temperature (1073 K), well above that expected for ≤20 K grain temperatures in the ISM (Draine 2003). Further, the length scale of the resulting stacking faults is a few rather than hundreds of nanometers, and these faults would not be expected to preferentially occur in one region of the grain. Thus, the observed stacking disorder in grains KH1, KH6,

381  and KH21 is not a radiation damage signature, but rather formed during condensation or subsequent
382  mechanical processing.

383  The laboratory-based analysis of ancient stardust has implications for remote astronomical
384  measurements. The composition and mineralogic makeup of planetary nebulae can be inferred based
385  on the comparison of spectra acquired from them with those acquired from laboratory standards. For
386  example, Hofmeister et al. (2004) compared the IR spectrum of the proto-planetary nebula NGC 6302
387  with silicates and several natural and synthetic materials in the $CaO-Al_2O_3$ system. They inferred that
388  forsterite and grossite are present in NGC 6302 and that hibonite most likely occurs there as well. The
389  grains that we report on here verify that RGB and AGB stars will condense single crystals of hibonite and
390  therefore provide corroborating evidence that such grains could be detectable within circumstellar or
391  nebular environments. We note, however, that the hibonites examined in this study deviate from the
392  pure $CaAl_{12}O_{19}$ used to match the IR spectra from NGC 6302, and the minor substitutional cations we
393  observe may affect the IR spectra. Future efforts aimed at modeling IR spectra from proto-planetary
394  nebulae and inferring their mineralogic makeup might consider incorporating meteoritic hibonites with
395  a range of compositions and grain shapes as reference standards.

396

397  **Summary and Conclusions**

398  We have reported the isotopic, crystal structure, and crystal chemistry data for five presolar
399  hibonite grains obtained from an acid residue of the Krymka LL3.1 ordinary chondrite. Isotopic
400  compositions indicate that: grains KH1 and KH15 formed in low-mass RGB/AGB stars undergoing cool-
401  bottom processing (deep mixing); KH6 and KH21 formed, respectively, in low-mass RGB/AGB (Z = ☉) and
402  AGB Z = (0.75☉) stars that evolved through first dredge-up; and grain KH2 condensed in the ejecta of a
403  Type II supernova.

The structural data reveals that all grains are single crystals with lattice parameters that are consistent with hibonite (S.G. $P6_3/mmc$; a = 0.556 nm, c = 2.19 nm). Of the five grains, only one (KH6) is pure $CaAl_{12}O_{19}$. All others contain variable amounts of Mg, Ti, Si, and Fe, but are otherwise stoichiometric. The single crystallinity of the grains and their stoichiometric compositions are generally consistent with predictions of equilibrium condensation models, which constrain the condensation temperatures to between 1480 K and 1743 K at total pressures of between $1 \times 10^{-3}$ and $1 \times 10^{-6}$ atm, respectively. The condensation conditions do not appear to vary significantly with the class of progenitor star, i.e., the condensation parameters for hibonite condensation around a supernova (grain KH2) are similar to those around AGB and RGB stars (KH1, KH6, KH15, and KH21). Aside from minor stacking disorder, we find no other significant structural perturbations in these five hibonite grains. In particular, consistent with previous studies of other presolar oxide grains, we do not observe any signatures of radiation processing. Any radiation damage that could be present occurs at a level that is below our imaging resolution (e.g. point defects).

Laboratory-based analysis of ancient stardust offers a measure of ground truth for observational astronomy. The grains we report on here unequivocally verify that stars evolving through the RGB and AGB stage of stellar evolution, as well as Type II supernovae, can condense single crystal stoichiometric hibonite grains. The complex morphology of grain KH15 and the observed deviations in minor element chemistry of four of the grains indicate that laboratory reference spectra from non-spherical hibonite with a range of compositions will be needed in order to model the infrared spectra acquired from circumstellar environments and proto-planetary nebulae.

**Acknowledgements**

We thank Dr. Tim McCoy (Smithsonian Institution) for the sample of Tanzania hibonite and Drs. Nabil Bassim, Ken Grabowski, and Graham Hubler (Naval Research Laboratory) for interesting discussions. We

428  also thank Dr. Christine Floss for a constructive review. This research was supported by the Office of
429  Naval Research and the NASA Cosmochemistry program (NNH09AL20I and NNX07AJ71G).
430
431

Table 1

Grain composition as measured using TEM-EDS and expressed in terms of wt% oxide and cation count.

| Grain | KH1 | KH2 | KH6* | KH15 | KH21 |
|---|---|---|---|---|---|
| CaO | 8.63 | 8.22 | 8.40 | 8.81 | 8.42 |
| $Al_2O_3$ | 87.76 | 89.58 | 91.60 | 88.72 | 89.18 |
| MgO | 0.81 | 0.86 | n.d. | 0.33 | 1.27 |
| $TiO_2$ | 1.36 | 1.06 | n.d. | 2.04 | 0.86 |
| $SiO_2$ | 0.78 | 0.18 | n.d. | n.d. | 0.13 |
| FeO | 0.67 | 0.10 | n.d. | 0.10 | 0.15 |
| Cations based on 19 O atoms | | | | | |
| Ca | 1.04 | 0.98 | 1.00 | 1.06 | 1.01 |
| Al | 11.58 | 11.77 | 12.00 | 11.69 | 11.73 |
| Ti | 0.11 | 0.09 | n.d. | 0.17 | 0.07 |
| Si | 0.09 | 0.02 | n.d. | n.d. | 0.01 |
| Mg | 0.14 | 0.14 | n.d. | 0.06 | 0.21 |
| Fe | 0.06 | 0.01 | n.d. | 0.01 | 0.01 |
| Group | 2 | 4 | 1 | 2 | 1 |
| Mass ($M_\odot$) | … | … | 1.4 | … | 1.5 |
| $^{17}O/^{16}O \pm 1\sigma$ | 6.59±0.11×10$^{-4}$ | 6.95±0.09×10$^{-4}$ | 5.77±0.08×10$^{-4}$ | 1.17±0.04×10$^{-3}$ | 6.84±0.16×10$^{-4}$ |
| $^{18}O/^{16}O \pm 1\sigma$ | 2.25±0.38×10$^{-4}$ | 4.78±0.04×10$^{-3}$ | 1.58±0.02×10$^{-3}$ | 4.66±0.17×10$^{-4}$ | 1.23±0.08×10$^{-3}$ |
| $\delta^{25}Mg/^{24}Mg \pm 1\sigma$ | … | -320±15 | … | -68±14 | -198±11 |
| $\delta^{26}Mg/^{24}Mg \pm 1\sigma$ | … | 7090±120 | … | 2882±55 | 13900±140 |
| $^{26}Al/^{27}Al \pm 1\sigma$ | … | 9.1±0.2×10$^{-3}$ | … | 8.2±0.2×10$^{-3}$ | 1.78±0.02×10$^{-2}$ |

n.d. = not detected

*nominal composition from standardless quantification

Table 2. Ti contents of presolar oxide grains. Isotopic data can be found in Nittler et al. (2008) except as indicated.

| Grain | Meteorite | Phase | Group | [Ti] (wt%)[a] |
|---|---|---|---|---|
| KH4 | Krymka | Hibonite | 1 | 0.39 |
| KH7 | Krymka | Hibonite | 1 | 0.56 |
| KH8 | Krymka | Hibonite | 1 | 0.61 |
| KH9 | Krymka | Hibonite | 1 | 0.59 |
| KH10 | Krymka | Hibonite | 1 | 0.34 |
| KH12 | Krymka | Hibonite | 1 | 0.48 |
| KH14 | Krymka | Hibonite | 1 | 0.58 |
| KH16 | Krymka | Hibonite | 1 | 0.14 |
| KH17 | Krymka | Hibonite | 1 | 0.52 |
| KH19 | Krymka | Hibonite | 1 | 0.40 |
| KH21 | Krymka | Hibonite | 1 | 0.52* |
| KH1 | Krymka | Hibonite | 2 | 0.82* |
| KH13 | Krymka | Hibonite | 2 | 1.03 |
| KH15 | Krymka | Hibonite | 2 | 1.22* |
| KH18 | Krymka | Hibonite | 2 | 0.66 |
| KH11 | Krymka | Hibonite | 3 | 0.59 |
| KH2 | Krymka | Hibonite | 4 | 0.64* |
| KC23 | Krymka | $Al_2O_3$ | 1 | 0.078 |
| T96[b] | Tieschitz | $Al_2O_3$ | 1 | <0.05* |
| T102 | Tieschitz | $Al_2O_3$ | 1 | 0.018 |
| T103[b] | Tieschitz | $Al_2O_3$ | 1 | 0.12 |
| T105 | Tieschitz | $Al_2O_3$ | 1 | 0.004 |
| T111 | Tieschitz | $Al_2O_3$ | 1 | 0.18 |
| KC26 | Krymka | $Al_2O_3$ | 1 | 0.12 |
| KC29 | Krymka | $Al_2O_3$ | 1 | 0.15 |
| T106 | Tieschitz | $Al_2O_3$ | 2 | 0.030 |
| T107 | Tieschitz | $Al_2O_3$ | 2 | 0.19 |
| KC25 | Krymka | $Al_2O_3$ | 2 | 0.32 |
| ORG114-12[c] | Orgueil | $Al_2O_3$ | 2 | 0.78 |
| KC30 | Krymka | $Al_2O_3$ | 2 | 0.11 |
| KC32 | Krymka | $Al_2O_3$ | 2 | 0.055 |

[a]Determined from NanoSIMS $^{48}Ti^+/^{27}Al^+$ ratios except * from TEM-EDS (this work)
[b]Stroud et al. (2004)
[c]Stroud et al. (2007)

**Figure Captions**

**Figure 1** Three-O-isotope plot for presolar oxide grains. Hibonite grains (shown as circles; filled indicates the grains in this study) are shown together with $Al_2O_3$ and $MgAl_2O_4$ grains (white and gray diamonds, respectively). Ellipses indicate the approximate location of the Groups into which the presolar oxides plot (see Nittler et al. 1997). Dashed lines indicate solar isotopic ratios. See Nittler et al. (2008) for data sources.

**Figure 2** Secondary electron images and NanoSIMS data of hibonite grains prior to FIB sectioning. **(a)** KH6, **(b)** NanoSIMS $\delta^{17}O/^{16}O$ ratio map of grain KH6, showing that the anomaly is limited to the hibonite (hb) grain sitting on top of a corundum (cor) grain, **(c)** KH21, **(d)** KH15, **(e)** KH1, and **(f)** KH2. All grains occur on a Au pedestal (black arrowhead with white outline) produced by sputtering during the SIMS analysis. Pt straps were deposited on the top surface of each grain in the FIB, except for KH21 on which we deposited C, to mitigate ion implantation and radiation damage, and transect along the line indicated by the white arrowheads.

**Figure 3** TEM data on grain KH6. (a) Bright-field TEM image. (b) SAED pattern. The shadow that extends diagonally (in b) from the bottom-right part of the image toward the center is the beam stop used to prevent the intense forward-scattered beam from saturating the image. The hibonite grain (Hb) occurs on top of isotopically normal (solar) corundum (Cor). Both the hibonite and corundum are sandwiched between the Pt strap and Au substrate.

**Figure 4** TEM data on grain KH21. **(a)** Bright-field image mosaic. The contrast variations that occur on the right side of the grain parallel to the bottom edge are due to stacking disorder. **(b)** STEM-HAADF image.

**(c)** SAED pattern. Diffuse streaking occurs along [0001]. **(d)** HRTEM image from right edge of grain revealing abundant stacking disorder.

**Figure 5** TEM data on grain KH15. **(a)** STEM-BF image. **(b)** STEM-HAADF image. **(c-e)** SAED patterns. The dashed circles **(a)** indicate the regions from which the SAED patterns **(c-e)** were acquired.

**Figure 6** TEM data on grain KH1. **(a)** Bright-field image mosaic. **(b)** STEM-HAADF image mosaic. **(c)** SAED pattern.

**Figure 7** TEM data on grain KH2. **(a)** Bright-field image. **(b)** STEM-HAADF image. **(c)** SAED pattern.

**Figure 8** Plot of Ti abundance (wt%) versus the $^{18}O/^{16}O$ ratio for hibonite and $Al_2O_3$ grains. The Ti abundance was calculated from SIMS data based on sensitivity factors derived from the TEM-EDS measurements. Presolar hibonites have higher Ti contents than presolar $Al_2O_3$ grains and $^{18}O$-depleted Group 2 grains appear to have higher Ti contents than Group 1 grains.

Fig. 1

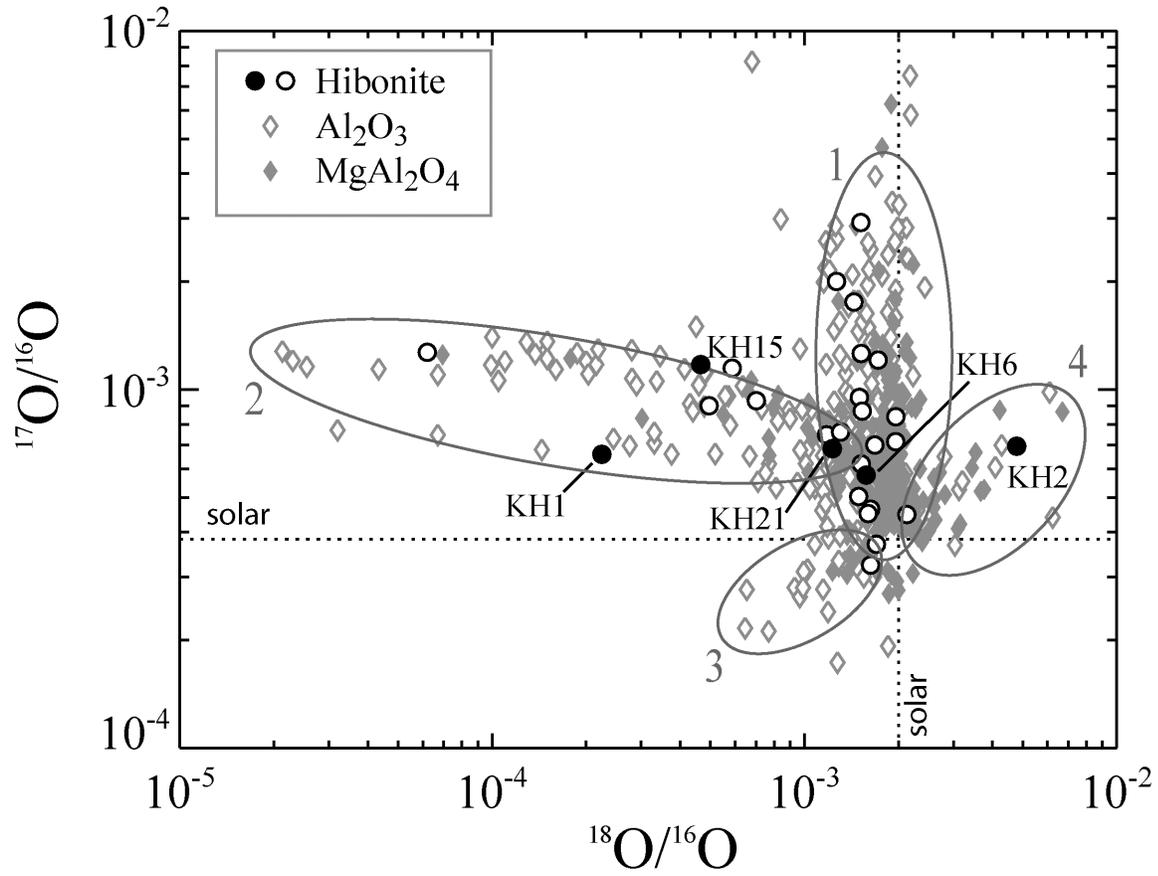

Fig. 2

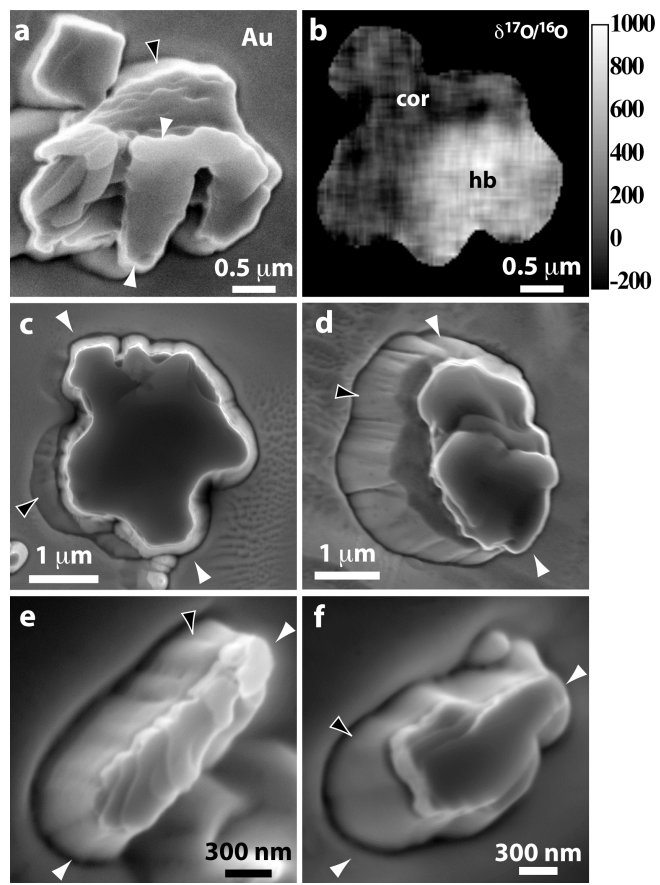

598   Fig. 3

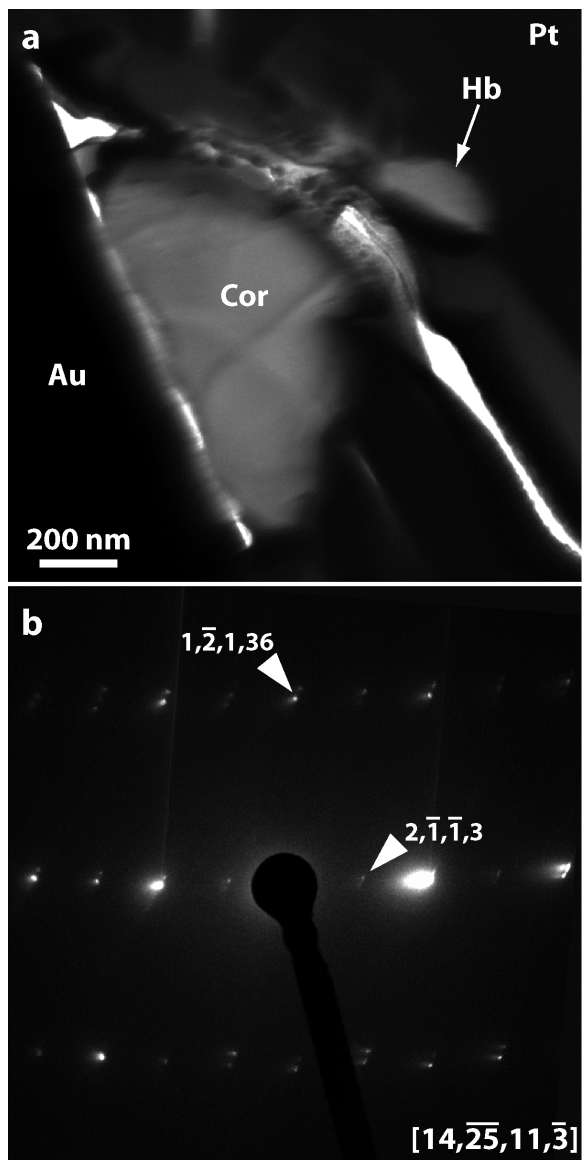

599
600
601

Fig. 4

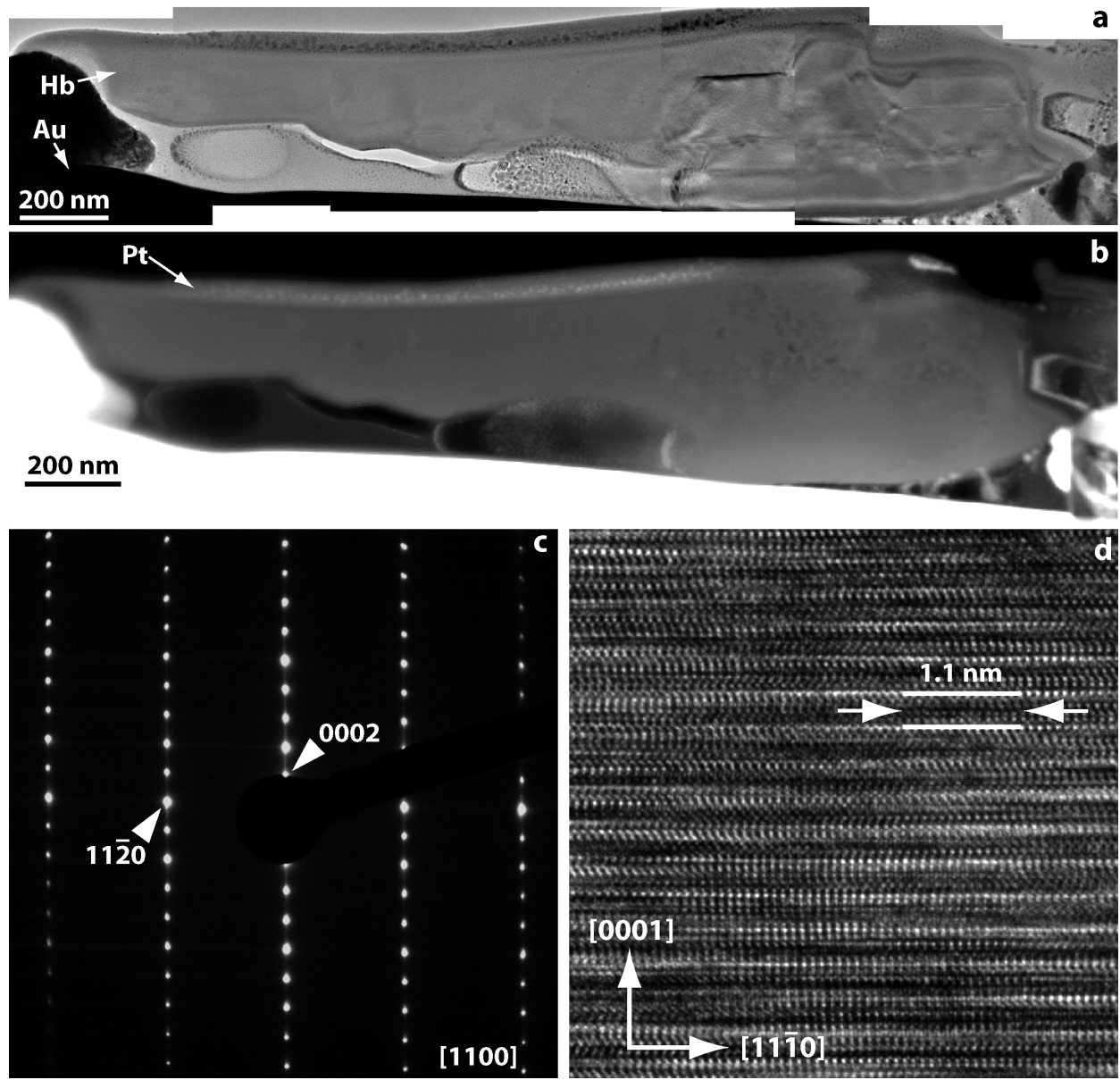

Fig. 5

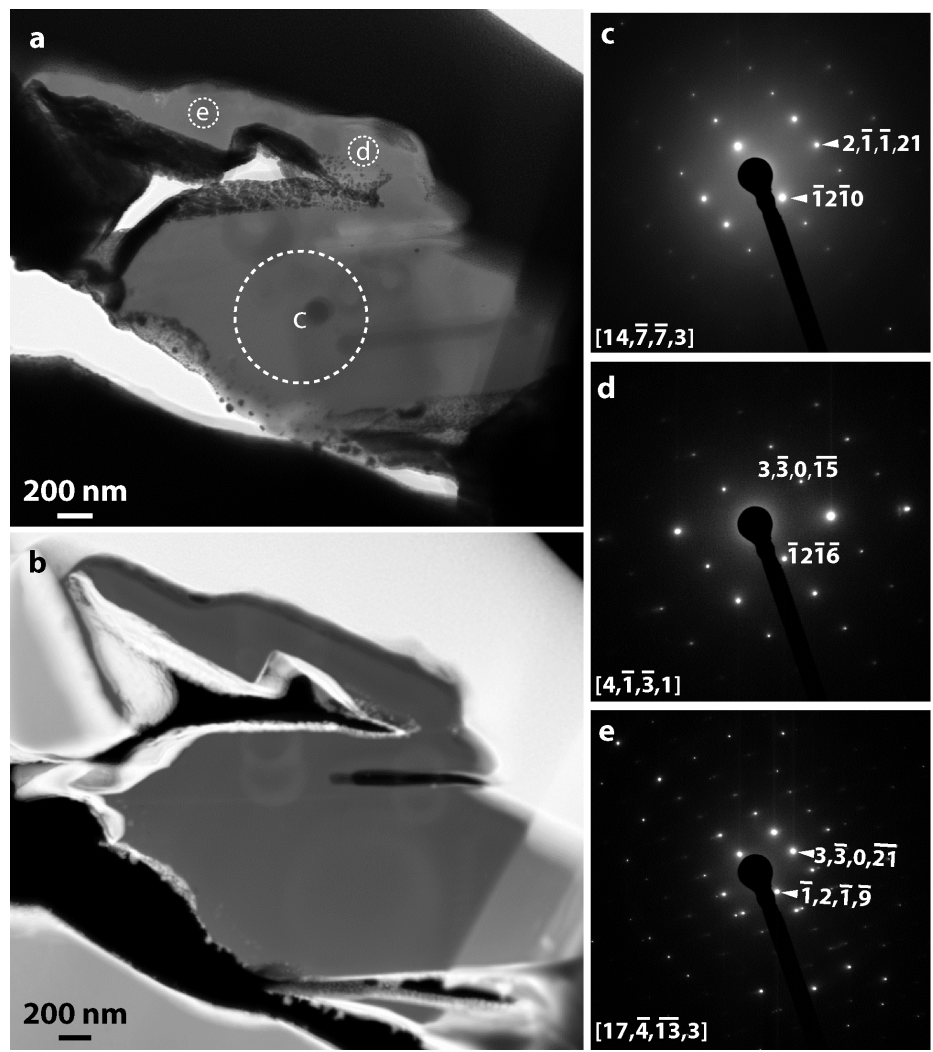

614

615 Fig. 6

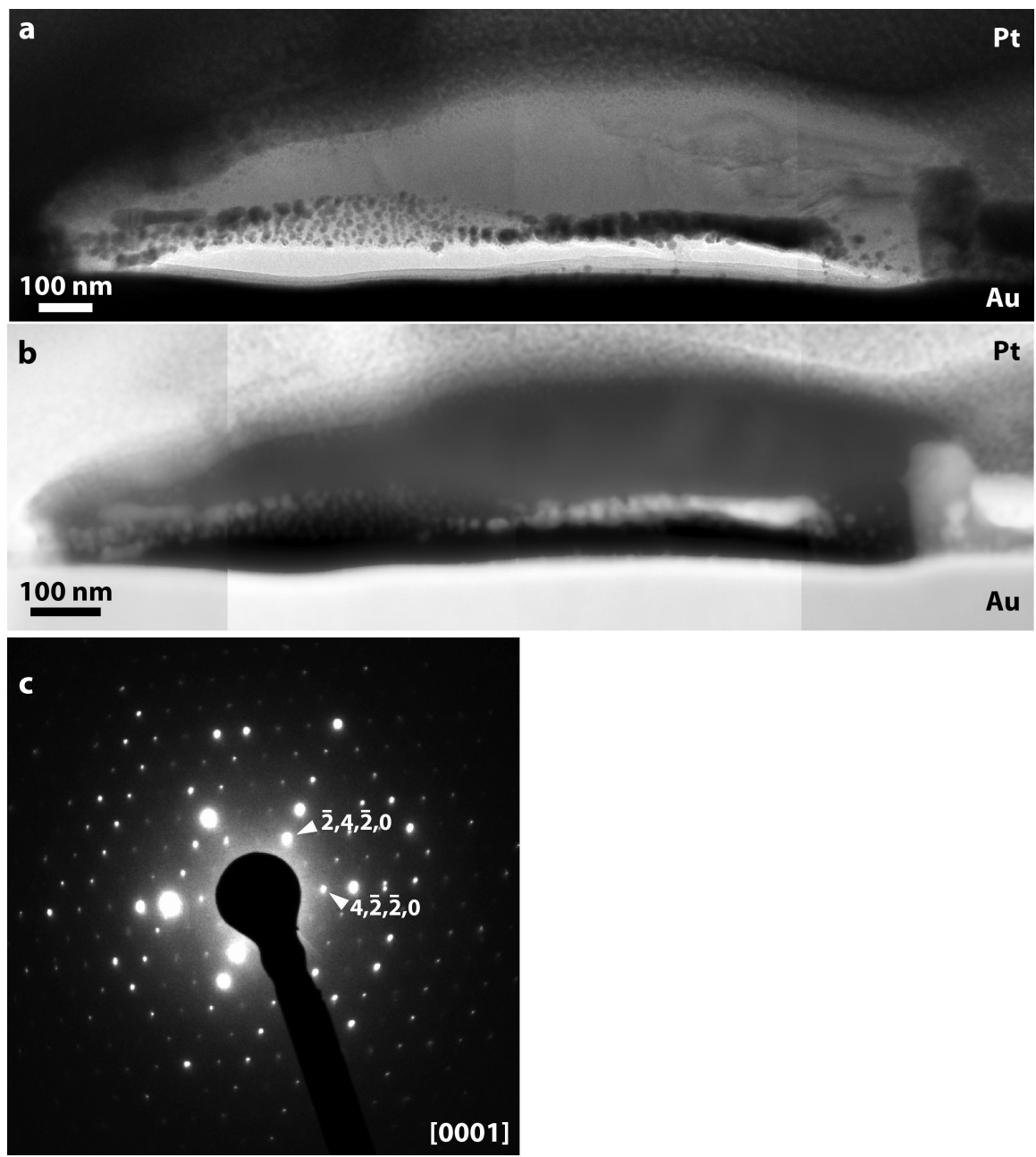

Fig. 7

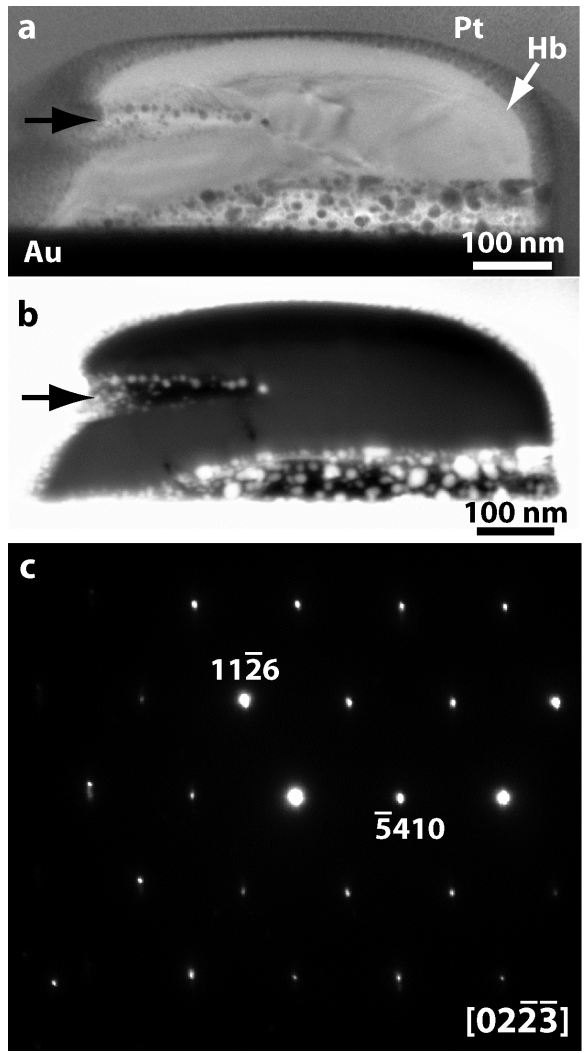

Figure 8

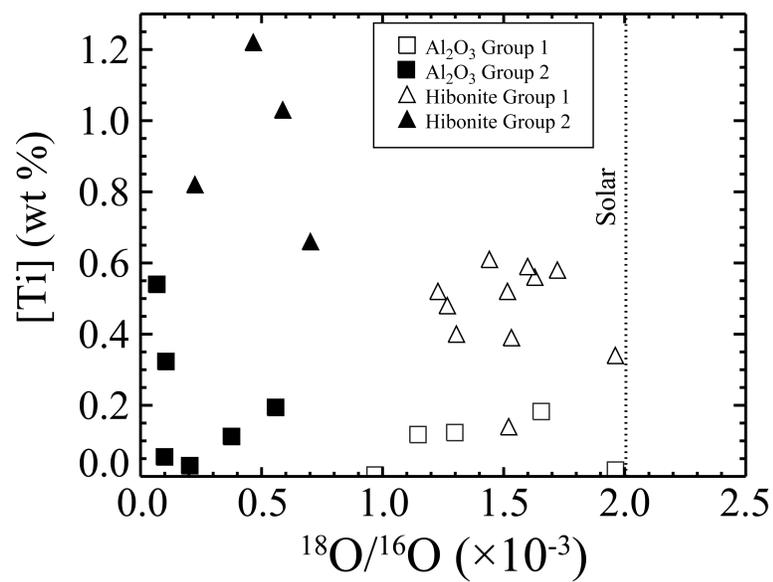